\documentclass[twocolumn,preprintnumbers,amsmath,amssymb,showpacs,prb,aps]{revtex4}
\usepackage{graphicx}
\usepackage{dcolumn}
\usepackage{amsmath}
\usepackage[perpage,symbol*]{footmisc}

\makeatletter
\def\btt#1{\texttt{\@backslashchar#1}}%
\DeclareRobustCommand\bblash{\btt{\@backslashchar}}%
\makeatother

\begin{document}

\title{Excitonic energy transfer in light-harvesting complexes in purple bacteria}

\author{Jun Ye $^{1}$, Kewei Sun$^{1}$, Yang Zhao$^{1}$\footnote{Electronic address:~\url{YZhao@ntu.edu.sg}}, Yunjin Yu $^{1,2}$, Chee Kong
Lee$^{1}$,  and Jianshu Cao$^3$}

\affiliation{ $^1$School of Materials Science and Engineering, Nanyang Technological University, Singapore 639798\\
                 $^2$College of Physics Science and Technology, Shenzhen University, Guangdong, China 518060 \\
                 $^3$Department of Chemistry, Massachusetts Institute of Technology, Cambridge, Massachusetts, USA 02139}
\date{\today}
\widetext

\begin{abstract}
Two distinct approaches, the Frenkel-Dirac time-dependent variation and the Haken-Strobl model, are adopted to study energy transfer dynamics in single-ring and double-ring light-harvesting systems in purple bacteria. It is found that inclusion of long-range dipolar interactions in the two methods results in significant increases in intra- or inter-ring exciton transfer efficiency. The dependence of exciton transfer efficiency on trapping positions on single rings of LH2 (B850) and LH1 is similar to that in toy models with nearest-neighbor coupling only. However, owing to the symmetry breaking caused by the dimerization of BChls and dipolar couplings, such dependence has been largely suppressed. In the studies of coupled-ring systems, both methods reveal interesting role of dipolar interaction in increasing energy transfer efficiency by introducing multiple intra/inter-ring transfer paths. Importantly, the time scale (~$4ps$) of inter-ring exciton transfer obtained from polaron dynamics is in good agreement with previous studies. In a double-ring LH2 system, dipole-induced symmetry breaking leads to global minima and local minima of the average trapping time when there is a finite value of non-zero dephasing rate, suggesting that environment plays a role in preserving quantum coherent energy transfer. In contrast, dephasing comes into play only when the perfect cylindrical symmetry in the hypothetic system is broken. This study has revealed that dipolar interaction between chromophores may play an important part in the high energy transfer efficiency in the LH2 system and many other natural photosynthetic systems.
\end{abstract}

\maketitle \narrowtext

\section{Introduction}

 An essential photochemical process in plants and photosynthetic bacteria is light-harvesting, in which solar photons are first absorbed by antenna complexes (also known as the antenna protein), and the resulting excitonic energy is transferred with a high efficiency to specialized pigment-protein complexes (PPC) or the reaction centers (RC). In a photosynthetic system of bacteriochlorophylls aggregates embedded in protein scaffolds, it has been suggested that the protein environment plays a critical role in protecting excitonic coherence, and consequently,  facilitating coherent energy transfer. The entire process of coherent energy transfer takes place in a very short period of time, from a few hundred femtoseconds to a few picoseconds. Recent studies have shown that the efficiency of energy transfer from  antenna pigments to the reaction centers is remarkably high (\emph{e.g.}, $>$95\%) \cite{Sauer, Blankenship, Grondelle}. Such a high efficiency of energy transfer has inspired the designs of efficient artificial systems to convert solar energy into other forms. However, an acute lack of reliable simulation techniques and theoretical tools for understanding the exact mechanisms of long time quantum coherence in such systems hinders the realistic application of the highly efficient energy transfer in designing artificial systems. Currently, the F\"orster resonance energy transfer theory (FRET) based on weak electronic coupling approximation and the Redfield equation based on Markovian approximation are the two commonly applied methods to deal with the energy transfer processes in these complexes. The classical F\"orster theory treats electronic coupling perturbatively and also assumes incoherent hopping of exciton between single donor and acceptor induced by point dipole-dipole interaction of transition dipoles\cite{Scholes} of the chromophores. Assumption of point dipole is inadequate in describing pigments' aggregation which results in the breakdown of the approximation \cite{Cheng}. To overcome the drawback of the point-dipole approximation, Jang \emph{et al.}\cite{Jang1,Jang2,Jang3,Jang4,Jang5} developed a muti-chromophore version of FRET theory which can be applied to cope with the transfer pathway interference that is not accounted for in the previous FRET treatments. However, when the electronic coupling is strong but the system-bath coupling is weak, it is necessary to consider relaxation between delocalized exciton states. In this limit, the dynamics of exciton energy transfer (EET) can be described by the coupled Redfield equations in the exciton basis \cite{Breuer, Redfield, Knoester, Schreiber}.  Redfield equation, on the other hand, is applicable in the weak exciton-phonon coupling limit where the problem can be treated perturbatively. More importantly, in the Redfield equation \cite{Coker}, the bath degree of freedom has been projected out to obtain a reduced master equation for the density matrix evolution under the Markovian approximation. However, for a functioning photosynthetic system, the exciton-phonon coupling is not weak and also Markovian approximation is a poor one in describing the realistic bath behavior since the typical phonon relaxation time scale is quite long.

In purple bacteria, the antenna complexes have highly symmetric multi-ring structures\cite{McDermott, Koepke}, which are the subject of numerous studies regarding the mechanism underlying the highly efficient energy transfer processes \cite{Brixner, Grondelle, Zhao, Alexandra, Fassioli, Muh, Ishizaki}. Photons are absorbed mainly by the peripheral antennae complex LH2 of purple bacteria, where the photo-generated excitons are subsequently transferred to the reaction center within the harvesting complexe LH1 in a few picoseconds \cite{Hu1}. The LH2 complex is an eight-unit circular aggregate built from $\alpha\beta$-heterodimers forming C8 symmetry \cite{McDermott, Koepke}. Each unit contains a pair of $\alpha$ and $\beta$ apoproteins, three bacteriochlorophylls-a(BChls-a) molecules and a carotenoid, and the BChl-a molecules form two rings named according to their corresponding absorption maxima at 800nm and 850nm as the B800 and B850 rings, respectively. The B850 ring consists of 16 tightly positioned BChls-a, with the Mg-Mg distance about 9.36 ${\rm \AA}$ for the $1\alpha-1\beta$ dimer, and about 8.78 ${\rm \AA}$ for the $1\alpha-1\beta$ dimer \cite{Zhao}. There are many studies on the mechanism behind these highly efficient photosynthetic processes \cite{Brixner, Grondelle, Zhao, Alexandra, Fassioli, Muh, Ishizaki}. In Ref.~\cite{Cao1}, optimal energy transfer conditions in linear and nonlinear systems were analyzed, and  a similar approach was recently applied to self-assembling light-harvesting arrays \cite{Cao2} to study optimal efficiency. For the antenna complexes of purple bacteria, energy transfer dynamics in LH1 and LH2 rings is inadequately understood, and questions remain on regarding how dipolar interaction, symmetry of the ring and inter-ring coupling etc., affect the energy transfer efficiency.

Inspired by previous studies, in this work we study in detail the effect of intra-ring dipolar interactions, ring geometry and symmetry, and more importantly, inter-ring distances in determining EET. As a starting point of understanding EET events, two approaches have been adopted to investigate various light-harvesting complexes in purple bacteria. For the low-temperature exciton dynamics, we use the Frenkel-Dirac time-dependent variational method to simulate the quantum dynamics of the Holstein polaron via the Davydov D1 Ansatz, while for the high-temperature dynamics and transfer efficiency calculations, the Haken-Strobl model \cite{Haken}, equivalent to the Redfield equation at high temperatures, has been employed.

This paper is organized as follows. In section II, the methodology of aiming for establishing possible optimal conditions and understanding of dynamics for energy transfer in the ring systems has been proposed. The effect of various parameters including the effect of dynamics disorder due to coupling to phonons and dephasing due to inclusion of environment in determining optimal energy transfer efficiency has been discussed in details in section III. Finally the conclusions are drawn in section IV.

\section{Methodology}

Two approaches are adopted to study the energy transfer process in the light-harvesting apparatus of the purple bacteria, the Frenkel-Dirac time-dependent variational method making use of the Holstein model which can be generally regarded as low-temperature cases, and the Haken-Strobl method applicable to high-temperature scenarios.

\subsection{Polaron Dynamics in LH2 Systems}

The Holstein Hamiltonian for the exciton-phonon system reads as~\cite{Holstein1959, Mahan}
\begin{equation}\label{H_tot}
\hat{H}=\hat{H}_{\rm ex}+\hat{H}_{\rm ph}+\hat{H}_{\rm ex-ph}
\end{equation}
with
\begin{equation} \label{H_ex}
\hat{H}_{\rm ex}=-J\sum_{n}\hat{a}_{n}^{\dagger}(\hat{a}_{n+1}+\hat{a}_{n-1}),
\end{equation}
\begin{equation}\label{H_ph}
\hat{H}_{\rm ph}=\sum_{q}\omega_{q}\hat{b}_{q}^{\dagger}\hat{b}_{q},
\end{equation}
\begin{equation}\label{H_diag}
\hat{H}_{\rm ex-ph} = -\frac{1}{\sqrt{N}}\sum_{n}\sum_{q}g_q\omega_{q}\hat{a}_{n}^{\dagger}\hat{a}_{n}\Big(\hat{b}_{q}e^{iqn}+{\rm H.c.}\Big),
\end{equation}
here $\hat{a}_{n}^{\dagger}$ ($\hat{a}_{n}$) is the creation (annihilation) operator for an exciton at the $n^{\rm th}$ site, and $J$ is the exciton transfer integral. $\hat{b}_{q}^{\dagger}$ ($\hat{b}_{q}$) is the creation (annihilation) operator of a phonon with momentum $q$ and frequency $\omega_{q}$, and the Planck's constant is set as $\hbar=1$. $\hat{H}_{\rm ex-ph}$ is the linear, diagonal exciton-phonon coupling Hamiltonian with $g_q$ as the coupling constant, with $\sum_q{g_q^2}\omega_q = S\omega_0$. Here $S$ is the well-known Huang-Rhys factor, and $\omega_0$ is the characteristic phonon frequency, which is set to unity in this paper for simplicity. It is noted that in this method, phonon dispersions in the Hamiltonian can have various forms to describe many phonon branches of different origins. For example, a linear phonon dispersion with $\omega_{q}=\omega_{0}[1 + W(2|q|/\pi-1)]$ may be adopted, where $W$ is a constant between 0 and 1, and the band width of the phonon frequency is $2W\omega_0$. Furthermore, other forms of exciton-phonon coupling can also be included, which implies the capability of current method in imitating the spectral density of phonon obtained with other methods such as molecular dynamics \cite{Ulrich}.

For multiple-ring system, by assuming only exciton transfer integral and dipolar interaction between rings, the original Holstein Hamiltonian can be modified as:
\begin{equation}\label{H_ex2}
{\hat{H}_{\rm ex}}^{'}=-\sum_{r_{1}r_{2}}\sum_{nm}J^{r_{1}r_{2}}_{nm}\hat{a}^{r_{1}\dagger}_{n}\hat{a}^{r_2}_{m},
\end{equation}
where $r_1=1...N_{\rm ring}$ and $r_2=1..N_{\rm ring}$ indicates any one pair of rings (with total number of $N_{\rm ring}$ rings) within the 2-dimensional  lattice if ${r_1}\neq{r_2}$, where the summation is over all rings. However if ${r_1}={r_2}$, Eq.~(\ref{H_ex2}) becomes:
\begin{equation}\label{H_ex3}
{\hat{H}_{\rm ex}}^{''}=-\sum_{r}\sum_{nm}J^r_{nm}\hat{a}^{r\dagger}_{n}\hat{a}^{r}_{m},
\end{equation}
where $r$ runs from 1 to $N_{\rm ring}$ rings. For each ring, the exciton Hamiltonian takes the form of Frenkel exciton Hamiltonian with $J_{nm}$ given by \cite{Hu2}:
\begin{equation}\label{J_nm}
J_{nm}=
\begin{pmatrix}
\varepsilon_{1} & J_{1} & W_{1,3} & \cdot & \cdot  & \cdot & J_{2}  \\
J_{1}& \varepsilon_{2} & J_{2}  & \cdot & \cdot & \cdot & W_{2,2N}  \\
W_{3,1} & J_{2} & \varepsilon_{1} & \cdot & \cdot  & \cdot & \cdot  \\
\cdot & \cdot  & \cdot  & \cdot  & \cdot  & \cdot  & \cdot  \\
\cdot  & \cdot  & \cdot  & \cdot &\varepsilon_{2} & J_{2} & W_{2N-2,2N} \\
\cdot  & \cdot  & \cdot  & \cdot& J_{2} &\varepsilon_{1}  & J_{1} \\
J_{2}  & \cdot  & \cdot  & \cdot & W_{2N,2N-2} & J_{1} &\varepsilon_{2}   \\
\end{pmatrix},
\end{equation}
where $\varepsilon_{1}$ and $\varepsilon_{2}$ are the on-site excitation energies of an individual BChl-a, $J_{1}$ and $J_{2}$ are the transfer integral between nearest neighbors, N equals 8 as the system is of C8 symmetry, and matrix elements $W_{i,j}$ are the dipolar coupling for the non-nearest neighbors. The dipolar coupling between site $i$ and site $j$ takes the form
\begin{equation}\label{dipole-dipole}
    W_{i,j}=C \Big[\frac{\textbf{d}_i \cdot \textbf{d}_j}{|\textbf{r}_{ij}|^3} - \frac{3( \textbf{d}_i \cdot\textbf{r}_{ij})(\textbf{d}_j\cdot\textbf{r}_{ij})}{|\textbf{r}_{ij}|^5}\Big],
\end{equation}
where C is the proportionality constant, $\textbf{r}_{ij}$ is the vector connecting the $i$th and $j$th monomers, and $\textbf{d}_i$ are the unit vectors of the $i$th BChl-a. In our calculations in this work, the following parameters\cite{Zhao} are adopted: $J_1=594\text{cm}^{-1}$, $J_2=491\text{cm}^{-1}$ and $C=640725{\rm \AA} ^3 \text{cm}^{-1}$. When the $J_nm$ matrix is applied to the polaron dynamics calculations, it must be scaled by the characteristic phonon frequency $\omega_0$. The Mg-Mg distance between neighboring B850 BChls is 9.2 ${\rm \AA}$ within the $\alpha \beta$-heterodimer and 9.25 ${\rm \AA}$ between neighboring heterodimers. For simplicity, the $\alpha$ and $\beta$ proteins are assumed to be evenly distributed with a distance of 9.25 ${\rm \AA}$.

For multiple-ring systems, the phonon and exciton-phonon interaction Hamiltonian can be described as:
\begin{equation}\label{H_ph2}
{\hat{H}_{\rm ph}}^{'}=\sum_r \sum_q \omega_q ^r \hat{b}_q ^{r\dagger} \hat{b}^r_q,
\end{equation}
\begin{equation}\label{H_exph2}
{\hat{H}_{\rm ex-ph}}^{'}=-\frac{1}{\sqrt{N}}\sum_r \sum_n \sum_q g^r_q\omega^r _q \hat{a}^{r\dagger}_n \hat{a}_n ^r (e^{iqn} \hat{b}^r_q+H_\cdot c_\cdot),
\end{equation}
where $r$ is summed over 1 to $N_{\rm ring}$. $\hat{b}_{q}^{r\dagger}$ ($\hat{b}^r_{q}$) is the creation (annihilation) operator of a phonon with momentum $q$ and frequency $\omega_{q}^r$ in the $r^{\rm th}$ ring. Similar to single-ring case, we can obtain the Huang-Rhys factor $S$ from $\sum_q({g_q^{r}})^2\omega_q^r = S\omega_0$. It is clearly shown in the formalism that the phonons in different LH2 rings are completely independent of each other in the current treatment.

The family of the time-dependent Davydov Ans\"{a}tze, i.e., the ${\rm D}_1$, $\tilde{\rm D}$ and ${\rm D}_2$ trial states, can be written in a general form as
\begin{equation}\label{localDavyAnsatz}
\big|\Psi_{\rm D}(t)\rangle = \sum_{n}\alpha_{n}(t)\hat{a}_{n}^{\dagger}\hat{U}_n^{\dagger}(t)|0\rangle_{\rm ex}|0\rangle_{\rm ph},
\end{equation}
where $\alpha_{n}(t)$ are the variational parameters representing exciton amplitudes, and $\hat{U}_n^{\dagger}(t)$ is the Glauber coherent operator
\begin{equation} \label{coherentOperator}
\hat{U}_n^{\dagger}(t) \equiv \exp\Big\{\sum_{q}\big[\lambda_{n,q}(t)\hat{b}_{q}^{\dagger}-{\rm H.c.}\big]\Big\}.
\end{equation}
For the ${\rm D}_1$ Ansatz, $\lambda_{n,q}(t)$ are $N{\times}N$ variational parameters representing phonon displacements. The ${\rm D}_2$ and $\tilde{\rm D}$ Ans\"{a}tze are two simplified cases of the ${\rm D}_1$ Ansatz. In the ${\rm D}_2$ Ansatz, $\lambda_{n,q}(t)$ are replaced by $N$ independent variational parameters $\beta_{q}(t)$, i.e., $\lambda_{n,q}(t) = \beta_{q}(t)$. While in the $\tilde{\rm D}$ Ansatz, $\lambda_{n,q}(t)$ are replaced by $2N-1$ variational parameters $\lambda_0(t)$, $\beta_{q\neq0}(t)$ and $\gamma_{q\neq0}(t)$, where $\lambda_{n,q=0}(t) = \lambda_0(t)$, and $\lambda_{n,q}(t)=\beta_{q}(t)+e^{-iqn}\gamma_{q}(t),~~q\neq0$.

The time evolution of the photo-excited state in a one-dimensional molecular aggregate follows the time-dependent Schr\"{o}dinger equation. There are several approaches to solve the time-dependent Schr\"{o}dinger equation. In the Hilbert space, for example, the time-dependent wave function $|\Phi(t)\rangle$ for the Hamiltonian $\hat{H}$ is parameterized by a set of time-dependent variables $\alpha_{m}(t)$ ($m=1,...,M$):
\begin{equation}
|\Phi(t)\rangle\equiv|\{\alpha_{m}(t)\}\rangle.
\end{equation}
Assuming that $|\Phi(t)\rangle$ satisfies the time-dependent Schr\"{o}dinger equation, one has
\begin{equation}
i\frac{\partial}{\partial t}|\Phi(t)\rangle=\hat{H}|\Phi(t)\rangle.
\label{TD-equation}
\end{equation}
Explicitly putting in the Hamiltonian $\hat{H}$ of Eq.~(\ref{H_tot}) and writing $\partial|\Phi(t)\rangle/\partial t$ in terms of $\alpha_{m}(t)$ and their time-derivatives $\dot{\alpha}_{m}(t)$ ($m=1,...,M$), one obtains
\begin{equation}
(i\frac{\partial}{\partial t}-\hat{H})| \Phi(t)\rangle = |\{\alpha_{m}(t)\},|\{\dot{\alpha}_{m}(t)\}\rangle= 0\label{TD-equation2}.
\end{equation}
Projecting Eq.~(\ref{TD-equation2}) onto $M$ different states $|\Psi_{m}\rangle$ ($m=1,...,M$), one obtains $M$ equations of motion for the parameters $\alpha_{m}(t)$:
\begin{equation}
\langle\Psi_{m}|\{\alpha_{m}(t)\},\{\dot{\alpha}_{m}(t)\}\rangle=0 \label {projection}.
\end{equation}
The approach we adopt in this work is the Lagrangian formalism of the Dirac-Frenkel time-dependent variational method \cite{Itzykson}, a powerful technique to obtain approximate dynamics of many-body quantum systems for which exact solutions often elude researchers. We formulate the Lagrangian $L$ as follows
\begin{equation}
 L=\langle\Phi(t)|{\frac{i\hbar}{2}}\frac{\overset{\longleftrightarrow}{\partial}}{\partial t}-\hat{H}|\Phi(t)\rangle.
\end{equation}
From this Lagrangian, equations of motion for the $M$ functions of time, parameters $\alpha_{m}(t)$, and their time-derivatives $\dot{\alpha}_{m}(t)$ ($m=1,...,M$), can be obtained by
\begin{equation}
\frac{d}{dt}(\frac{\partial
L}{\partial\dot{\alpha_{m}^{*}}})-\frac{\partial
L}{\partial\alpha_{m}^{*}}=0. \label {Lagrangian}
\end{equation}
Furthermore, to better elucidate the correlation between the exciton and the phonons, the phonon displacement $\xi_{\rm n}(t)$ is introduced and defined as
\begin{equation}\label{xn}
\xi_n(t) \equiv \Big\langle\Psi_{\rm D}(t)\Big|\frac{\hat{b}_n+\hat{b}_n^{\dagger}}{2}\Big|\Psi_{\rm D}(t)\Big\rangle,
\end{equation}
where $\hat{b}_n^{\dagger}=N^{1/2}\sum_q e^{-iqn}\hat{b}_q^{\dagger}$. The reader is referred to Appendix A for discussions on the precision of the Davydov trial states (defined as $\Delta(t)$) and Appendix B for detailed derivations of equations of motion for polaron dynamics in a multiple-ring system.

\subsection{The Haken-Strobl model}

The second approach applied to the LH1/LH2 complexes in this work is the Haken-Strobl model, which has been previously used to describe the exciton dynamics in FMO \cite{Cao1}. We start with the Liouville equation  for the exciton density matrix $\rho$
\begin{align}\label{eq1}
\dot{\rho}(t) &=-\mathcal{L}\rho(t) \notag \\
 &=-[\mathcal{L}_{\rm sys}+\mathcal{L}_{\rm dissip}+\mathcal{L}_{\rm decay}+\mathcal{L}_{\rm trap}]\rho(t).
\end{align}
Here $\mathcal{L}_{\rm sys}$ is the free evolution operation of the pure exciton system, and can be written as $\mathcal{L}_{\rm sys}=i[H,\rho]/\hbar$, where $[H]_{nm}=(1-\delta_{nm})J_{nm}+\delta_{nm}\epsilon_{n}$ with $J_{nm}$ the excitonic coupling between site $n$ and site $m$, and $\epsilon$ the site energy. $\mathcal{L}_{\rm dissip}$ represents the dephasing and population effects within the exciton manifold. For the convenience of analysis, the coupling to the environment is given by the standard Bloch-Redfield equation, which can be written as $[\mathcal{L}_{\rm dissip}]_{nm}=(1-\delta_{nm}\Gamma_{nm})$, where $\Gamma$ is the pure dephasing rate. $\mathcal{L}_{\rm decay}$ represents the decay of the exciton to the ground state and can be expressed as $[\mathcal{L}_{\rm decay}]_{nm}=(k_{d,n}+k_{d,m})/2$ with $k_{d,n}$ the decay rate on site $n$. $\mathcal{L}_{\rm trap}$ represents the trapping rate, which is critical to the energy absorbing process, and can be described as $[\mathcal{L}_{\rm trap}]_{nm}=(k_{t,n}+k_{t,m})/2$ with $k_{t,n}$ the trapping rate on site $n$.

In fact, among $\mathcal{L}_{\rm decay}$ and $\mathcal{L}_{\rm trap}$, which represent two possible channels for irreversible exciton energy loss, the former is ineffective in the energy transfer. Efficiency of energy transfer can be gauged by the quantum yield $q$, namely the trapping probability~\cite{Ishizaki},
\begin{equation}\label{eq3}
q=\frac{\sum_{n}k_{t,n}\tau_{n}}{\sum_{n}k_{t,n}\tau_{n}+\sum_{n}k_{d,n}\tau_{n}},
\end{equation}
where the mean residence time $\tau_{n}$ can be expressed as $\tau_{n}=\int_{0}^{\infty}\rho_{n}(t)dt$, and the population $\rho_{n}=\rho_{nn}$ is the diagonal element of the density matrix. In photosynthetic systems, $k_{t}^{-1}$ is on the order of ${\rm ps}$ and $k_{d}^{-1}$ is on the order of $\rm ns$, so the trapping rate is much larger than decay rate, and the quantum yield is close to unity. The quantum yield can then be approximated as $q\thickapprox({1+k_{d} \langle t \rangle})^{-1}$, where $\langle t \rangle =\sum_n \tau_n$ is the mean first passage time to the trap state without the presence of the constant decay, \emph{i.e.,} the average trapping time. Quantum yield and trapping time have been extensively studied in the context of molecule photon statistics~\cite{Cao1} and experimental investigation was also carried out in a photosynthetic system \cite{Hofmanm}. A shorter trapping time corresponds to a higher efficiency of exciton transfer.

\section{Results and Discussions}

\subsection{Polaron Dynamics in LH2 Systems}

\begin{figure}
\begin{center}
\includegraphics[scale=0.33]{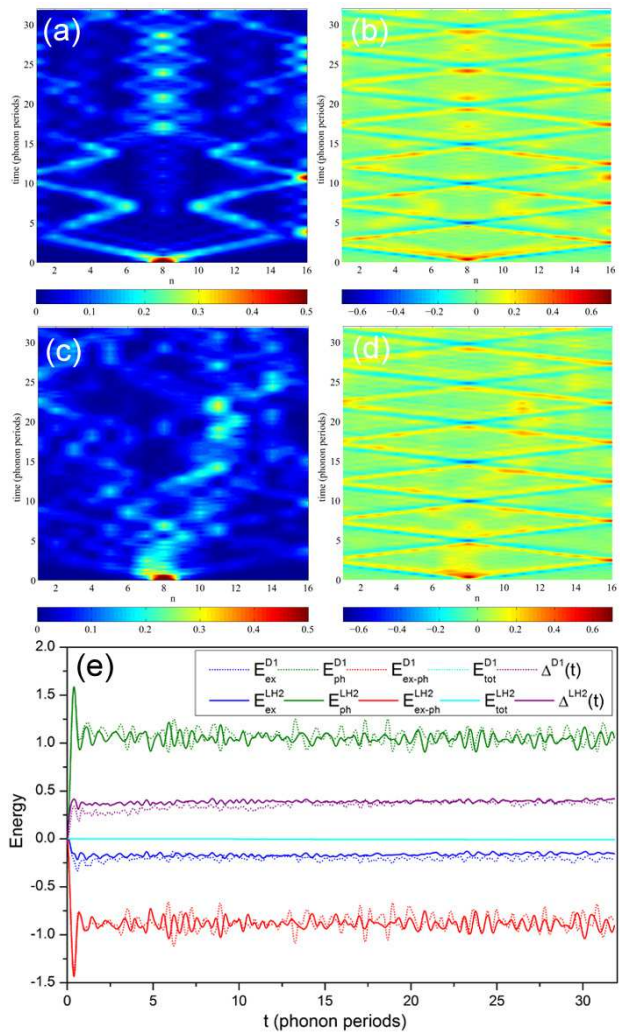}
\caption{Real space dynamics of (a) exciton probability $|\alpha_n(t)|^2$ and (b) phonon displacement $\xi_n(t)$ for a single-ring with only nearest neighbor transfer integral J = 0.3557, a moderate exciton-phonon coupling strength S = 0.5, and a wide bandwidth W = 0.8 of phonon dispersion. Corresponding real space dynamics are displayed for (c) exciton probability $|\alpha_n(t)|^2$ and (d) phonon displacement $\xi_n(t)$ for a single realistic LH2 ring transfer integral matrix. The D1 Ansatz is used for the simulation. The time evolution of energy components for exciton, phonon, exciton-phonon interaction and deviation magnitude $\Delta(t)$ for both cases has also been given in (e).  Note that the value of the nearest neighbor transfer integral is taken from the largest element (scaled by the characteristic phonon frequency $\omega_0$) of the realistic LH2 transfer integral matrix.} \label{Fig1}
\end{center}
\end{figure}

With the aid of Frenkel-Dirac time-dependent variational method, a hierarchy of variational wave functions has been previously formulated to faithfully describe the dynamics of the Holstein polaron~\cite{pccp, pssc}. The accuracy of the Davydov Ans\"atze has been discussed in length in Ref.~\cite{pccp, pssc}, and it has been shown that for diagonal exciton-phonon coupling, the D1 Ansatz has the highest precision in the hierarchy of the Davydov trial states when applied to the solution of Holstein Hamiltonian. The increase in the precision of the D1 Ansatz is achieved mainly by improved sophistication of the phonon wave function~\cite{pccp, pssc}. Such increased sophistication plays a dramatic role especially in the weak and moderate coupling regime, where the plane-wave-like phonon components render impractical to use a simple linear superposition of phonon coherent states. The vastly increased number of variational parameters for the phonon wave function ultimately results in higher precision of the D1 Ansatz.

The approach is first to be applied to the exciton polaron evolving in an isolated LH2 ring. To assure the accuracy of our approach, the D1 Ansatz has been chosen to deal with the dynamics in this paper. For a case of wide phonon bandwidth (W=0.8) and a moderate exciton-phonon coupling strength (S=0.5), time evolution of exciton probability $|\alpha_n(t)|^2$ and phonon displacement $\xi_n(t)$ for the case with only nearest neighbor transfer integral have been displayed in Figs.~\ref{Fig1}(a) and (b). The polaron state is initialized as follows: $\alpha_n(0)=\delta_{n,8}$ and $\lambda_{n,q}(0)=0$. The corresponding dynamics with realistic LH2 transfer integral matrix (\emph{cf.,} Eq.~(\ref{J_nm})) has been given in Figs.~\ref{Fig1}(c) and (d). Time-dependent exciton and phonon energies as well as the magnitude of deviation vector $\Delta(t)$ for both cases are plotted in Fig.~\ref{Fig1}(e). For all calculations, a characteristic phonon frequency of $\omega_0=1670 {\rm cm}^{-1}$ and a Huang-Rhys factor of $S$=0.5 are chosen\cite{Schulten1}. The phonon frequency obtained from molecular dynamics simulation is related to the stretching mode in either a C=O bond or a methine bridge.

\begin{figure}
\begin{center}
\includegraphics[scale=0.3]{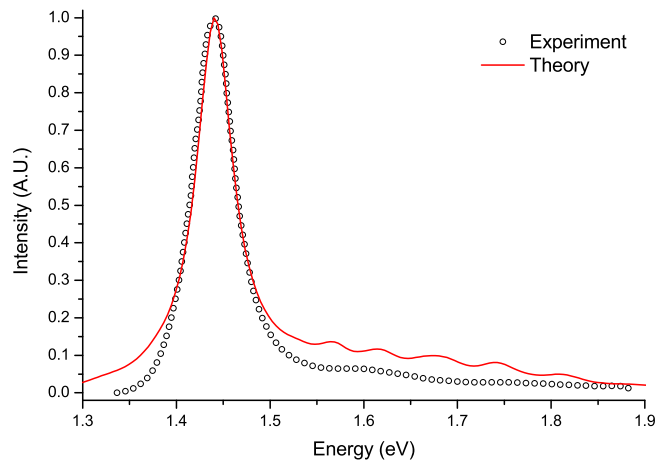}
\caption{Comparison of calculated linear absorption spectra and measured ones. A single-ring with realistic LH2 ring (B850) transfer integral matrix is adopted. Parameters used are S = 0.5, $\omega_0 $= 1670${\rm cm}^{-1}$, W = 0.8 and a damping factor of 0.12.} \label{Fig2}
\end{center}
\end{figure}

It is important to compare calculated absorption spectra with measured ones in order to further validate our approach and lend support to our simulation of excitation transfer dynamics in a photosynthetic complex. Linear absorption spectra are calculated, and compared to measured ones in Fig.~\ref{Fig2}. Detailed derivations for absorption spectral calculations can be found in Ref.~\cite{pccp,pssc} Measured data for B850 complexes are obtained from Ref.~\cite{absexp} Good agreement is found between theory and experiment, pointing to the applicability of our model to the B850 ring for the parameter set chosen. It is especially worth noting that the long tail shown in the measured spectrum is reproduced by our approach.

Owing to the aforementioned higher accuracy of the D1 Ansatz as recorded in Ref.~\cite{pccp, pssc} in comparison with other trial states, as shown in Fig.~\ref{Fig1}(e), it is therefore an ideal candidate for the study of excitation dynamics of the LH2 system. The real LH2 system contains pigments that interact via long-range dipolar couplings as given by the pure exciton Hamiltonian with transfer integral matrix given in Eq.~(\ref{J_nm}). By introducing dipole-dipole interaction into the Hamiltonian, the accuracy for the D1 Ansatz has to be rechecked and compared to that for a model system with nearest neighbor coupling only. A first step toward the accuracy checkup is the comparison of polaron dynamics in Figs.~\ref{Fig1}(a), (b), (c) and (d), where dramatic changes can be found as a result of dipolar interactions.

In addition to the case with only nearest neighbor $J$ shown in Figs.~\ref{Fig1}(a) and (b), many new features have emerged as shown in Figs.~\ref{Fig1}(c) and (d). One major finding is that the inclusion of realistic dipolar interactions in the LH2 system breaks the symmetry of the ring structure, resulting in asymmetric and faster exciton dynamics compared to the model system with only nearest neighbor $J$. Such an effect can be related to an emergence of additional energy transfer paths when dipolar interactions are included. Although the dynamics here does not contain the coupling to the external heat bath, the essential physics is captured to a great extent. It is worth noting that in Figs.~\ref{Fig1}(b) and~\ref{Fig1}(d), time evolution of phonon displacement components, \emph{i.e.,} $\xi_n(t)$ with ``V"-shaped feature survived as the exciton Hamiltonian has been completely changed from nearest neighbor $J$ to a realistic $J_{nm}$. Which can be explained as follows: the phonon dynamics of the system is not only determined by the coupling between exciton and phonon, but also has its own propagation pattern which is determined by the linear dispersion relationship given earlier, thus once the excitation on site 8 starts to propagate as shown in the exciton dynamics, the strong lattice distortion in the vicinity of the excitation will also propagate. Moreover, the effect of the exciton on phonon dynamics is apparent by comparing Figs.~\ref{Fig1}(a) and (b), where an exact correspondence between the exciton dynamics and phonon can be found. However, the value of exciton-phonon coupling strength $S$ tells us the coupling between exciton and phonon is moderate, thus the exciton generated on site 8 can easily propagate through larger distance similar to a bare exciton.

\begin{figure}
\begin{center}
\includegraphics[scale=0.33]{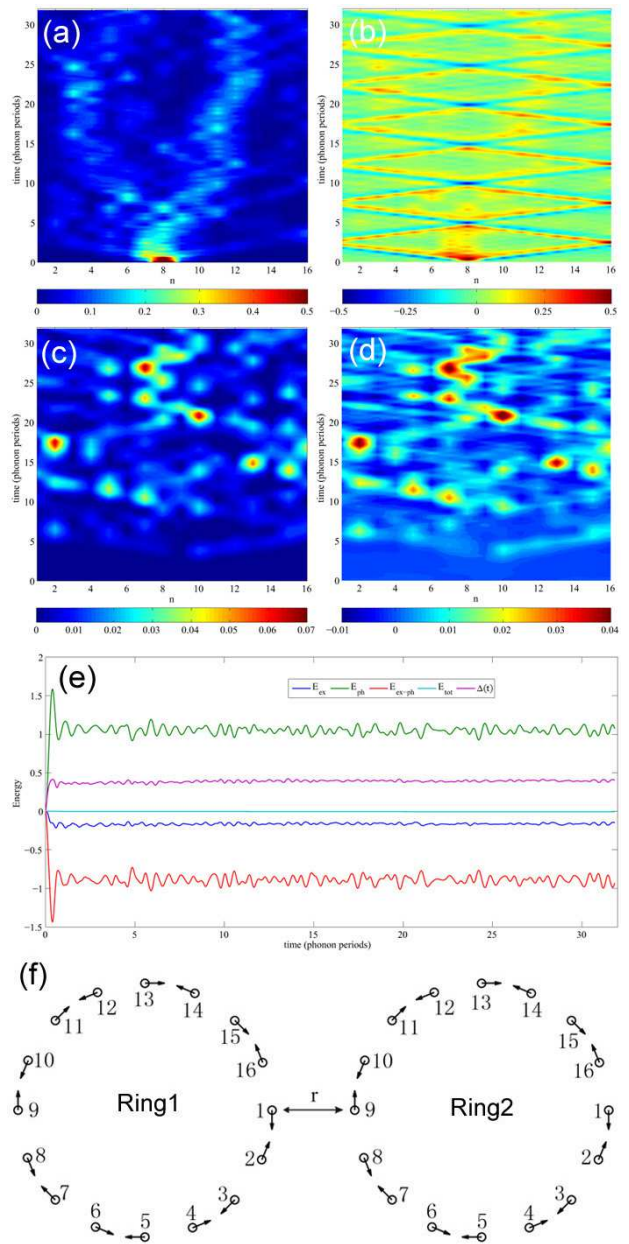}
\caption{Real space dynamics of (a) exciton probability $|\alpha_n(t)|^2$ and (b) phonon displacement $\xi_n(t)$ in ring-1 for a two-LH2-ring system with realistic LH2 ring transfer integral matrix, a moderate exciton-phonon coupling strength S = 0.5 and a wide bandwidth W = 0.8 of phonon dispersion. The D1 Ansatz is used for the simulation. Corresponding real space dynamics in ring-2 are displayed for (c) exciton probability $|\alpha_n(t)|^2$ and (d) phonon displacement $\xi_n(t)$. The time evolution of energy components from exciton, phonon, exciton-phonon interaction, deviation magnitude $\Delta(t)$ has also been given in (e). The distance between two rings is set to \textbf{15} ${\rm \AA}$, with interaction between two rings given by Eq.~(\ref{dipole-dipole}). (f) The schematics of two connected LH2 complexes. $r$ is the distance between the tips of the rings.} \label{Fig3}
\end{center}
\end{figure}

\begin{figure}
\begin{center}
\includegraphics[scale=0.33]{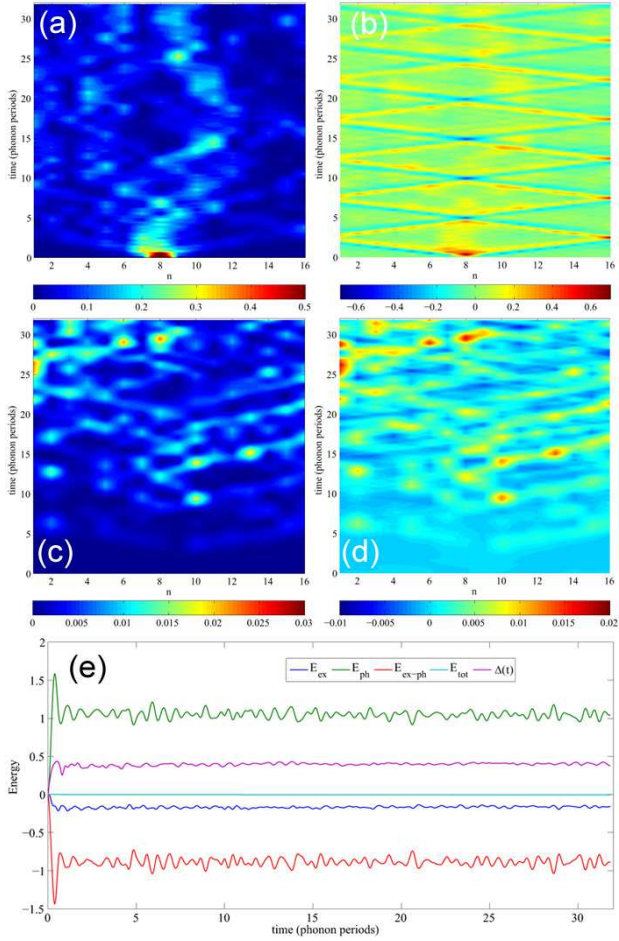}
\caption{Real space dynamics of (a) exciton probability $|\alpha_n(t)|^2$ and (b) phonon displacement $\xi_n(t)$ in ring-1 for a two-LH2-ring system with realistic LH2 ring transfer integral matrix, a moderate exciton-phonon coupling strength S = 0.5 and a wide bandwidth W = 0.8 of phonon dispersion. The D1 Ansatz is used for the simulation. Corresponding real space dynamics in ring-2 are displayed for (c) exciton probability $|\alpha_n(t)|^2$ and (d) phonon displacement $\xi_n(t)$. The time evolution of energy components from exciton, phonon, exciton-phonon interaction, deviation magnitude $\Delta(t)$ has also been given in (e). The distance between two rings is set to \textbf{20} ${\rm \AA}$, with interaction between two rings given by Eq.~(\ref{dipole-dipole}).} \label{Fig4}
\end{center}
\end{figure}

More interestingly, the revival and localization phenomenon shown in Fig.~\ref{Fig1}(a) is nearly absent in Fig.~\ref{Fig1}(c) as a result of broken symmetry as well as multiple transfer channels due to the presence of dipolar interaction within the ring, although there are certain possible positions such as on site 8 at around 6 phonon periods appears to have possible revival of exciton amplitude similar to those shown in Fig.~\ref{Fig1}(a) on site 8 at around 10 phonon periods, lending further support to the speed up of exciton dynamics (in terms of delocalization) due to the inclusion of dipolar interaction. From above discussions, by including long-range excitonic coupling other than nearest neighbor (which is true for many realistic cases such as the LH2 system investigated in this paper), the exciton can more efficiently and evenly spread over the entire ring in the presence of exciton-phonon coupling. In the other words, nature has designed a unique way for the light harvesting complexes such as LH2 in maximizing their excitonic energy transfer efficiency in spite of exciton-phonon coupling responsible for localization and dephasing of the excitons. It appears the symmetry breaking of the energy transfer pathway has more positive role in LH2 complexes.

Furthermore, It is also important to note in Fig.~\ref{Fig1}(e) that for the chosen parameter sets, the D1 Ansatz is capable to describe the realistic LH2 system with great precision. Comparing to the maximum absolute value of phonon or exciton-phonon interaction energy shown in the figure, the magnitude of deviation vector $\Delta(t)$ is less then 30 percent of the average system energy (here we use phonon energy as the reference). Inclusion of complex dipolar interaction does not further increase the error of the D1 Ansatz in describing the system dynamics. Although the deviation magnitude does not change much, the dynamics of exciton and phonon deviate dramatically if one compares Figs.~\ref{Fig1} (a), (b) and~\ref{Fig1} (c), (d).

Using the D1 Ansatz, we also examined exciton-phonon dynamics of a realistic two-ring system with long-range dipolar interactions. Results from the dynamics calculations are shown in Fig.~\ref{Fig3} and Fig.~\ref{Fig4} for two inter-ring distances (15 and 20 $\rm \AA$, respectively). The choice of the distances for the two-ring system is in accordance with the one shown in Ref.~\cite{Schulten1}, which suggests the Mg-Mg distance of around 24 $\rm \AA$ for the closest chromophores between two adjacent LH2 rings. Considering the size of the chromophores, it is beneficial to slightly reduce the distance to consider the exciton transfer from the tip of one molecule to the other, thus two different distances, \emph{i.e.,} 15 and 20 $\rm \AA$ are used in this study.  To have better understanding of the roles of intra- and inter-ring excitonic transfer, the two-ring system is initialized as follows: $\alpha^r_n(0)=\delta_{r,1}\delta_{n,8}$ and $\lambda^r_{n,q}(0)=0$, \emph{i.e.,} a single excitation at site 8 of ring-1, after a subsequential spreading within ring-1, the exciton population is transferred to site 1, which is the closet point to site 10 (and nearby sites) of ring-2, at about 3 phonon periods from t=0. This exciton transfer process is clearly revealed in Figs.~\ref{Fig3}(c) and~\ref{Fig4}(c).

It is always helpful to compare the value of $\Delta(t)$ relative with system energies as the first check. As shown in Figs.~\ref{Fig3}(e) and~\ref{Fig4}(e), the accuracy of the D1 Ansatz applied to the two-ring system with different inter-ring distances does not alter much as compared to the one-ring system since the magnitude $\Delta(t)$ remains close to that for the one-ring case. The value of $\Delta(t)$ also becomes fairly stable after initial 5 phonon periods, therefore confirms that the method applied to study the multiple-ring exciton dynamics using time-dependent variational Ansatz is valid.

At a first glimpse, one may spot few differences between Fig.~\ref{Fig3} and Fig.~\ref{Fig4} during the first 10 phonon periods. One can also find resemblance of the exciton probabilities and phonon displacement between the two-ring system and the one-ring system in Fig.~\ref{Fig1} if comparison is made during the same period of time. This is due to the fact that a very small amount of exciton population is transferred to the second ring of the two-ring system during this period as clearly revealed in Figs.~\ref{Fig3}(c) and~\ref{Fig4}(c), thus the dynamics of both the exciton and phonons are essentially those of the one-ring case. However, after 10 phonon periods, as the transfer of exciton population increases, $|\alpha_n(t)|^2$  and $\xi_{n}(t)$ become quite different for the two cases of different inter-ring distances.

Interestingly, as one may observe from Figs.~\ref{Fig3}(c) and~\ref{Fig4}(c), many bright spots appear in the exciton probability contour of ring-2, indicating possible pathway interference within the ring. The origins of the bright spots are explained as follows. Among the nearest neighbors of ring-1 [cf.~Fig.~\ref{Fig3}(f)], sites 8 to 10 of the second ring show more pronounced changes in exciton probability, if one compares Figs.~\ref{Fig3}(c) and~\ref{Fig4}(c). It is also true that interactions between sites 1, 2 and 16 of ring-1 and sites 8, 9 and 10 of ring-2 dominate inter-ring transfer according to Eq.~(\ref{J_nm}), resulting in multiple inter-ring exciton transfer pathways.

Due to strong interactions between site 1 of ring-1 and site 9 of ring-2 owing to their proximity, it is somewhat counterintuitive to observe the relatively smaller exciton probability on site 9 of ring-2. The explanation lies in the dimerization of LH2 rings which results in opposite signs of dipole moments of site 1 of ring-1 and site 9 of ring-2, as shown in Fig.~\ref{Fig3}(f). Besides dimerization, it is also helpful to consider the intra-ring interference in ring-2 between exciton wave packets that travel through multiple paths from ring-1 to ring-2. Figs.~\ref{Fig3}(c) and~\ref{Fig4}(c) suggest that sites 8 and 10 of ring-2 (especially, site 10) can be treated as input sites strongly connected to ring-1. Appearance of bright spots on sites 8 and 10 can be simply understood as a combined effect of exciton transfer from ring-1 to ring-2 and constructive interference of exciton wave packets in ring-2. However, on locations far from site 10, such as sites 2, 5, 13 and 15 of the second ring, the effect of constructive interference is more prominent. As sites 2 and 16 of ring-2 are opposite to the aforementioned input sites, it is not surprising to observe bright spots on site 2 and 16 as a result of constructive interference, as shown in Figs.~\ref{Fig3}(c) and~\ref{Fig4}(c). Effects of multiple pathways for inter-ring exciton transfer may play a role in the appearance of additional bright spots on ring-2.

\begin{figure}
\begin{center}
\includegraphics[scale=0.3]{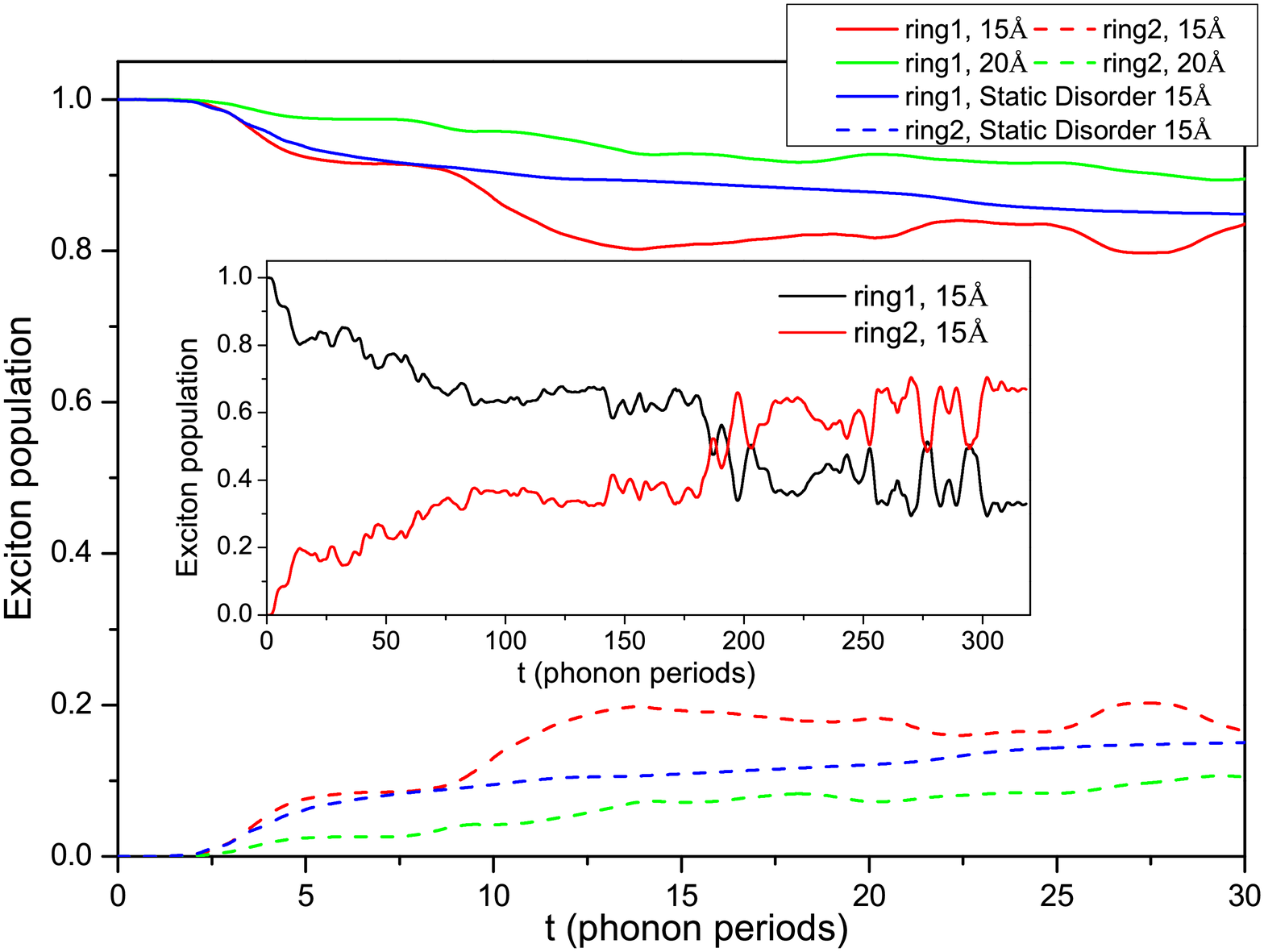}
\caption{Exciton population dynamics ($\sum_{n}|\alpha_{n}^r(t)|^2$) for the two different separations shown in Figs.~\ref{Fig2} and~\ref{Fig3}, where a long time dynamics of the exciton population has also been given in the inset for the ring system with a distance of 15  ${\rm \AA}$. Exciton population dynamics for the inter-ring distance of 15  ${\rm \AA}$ with on-site energy static disorder has also been given in the figure.}
\label{Fig5}
\end{center}
\end{figure}

Even more interestingly, one can also calculate the exciton population of a given ring, \emph{i.e.,} $\sum_n|\alpha_n^r (t)|^2$ in ring-$r$ (with $r$ the ring index) as a function of time to shed light on the overall exciton dynamics of the two-ring system with inter-ring distance of 15 and 20 $\rm \AA$. Calculation results are displayed in Fig.~\ref{Fig5}, where exciton populations corresponding to Figs.~\ref{Fig3} and~\ref{Fig4} are plotted as a function of time. A supplementary calculation for a much longer time is also given in the inset of Fig.~\ref{Fig5} for inter-ring distance of 15 $\rm \AA$. Short-time results suggest a nearly stepwise population transfer from ring-1 to ring-2. The stepwise features of population profiles in Fig.~\ref{Fig5} coincide with the increases of exciton probabilities on sites 8 to 10 in the second ring as shown in Fig.~\ref{Fig3}(c) and~\ref{Fig4}(c), lending further support to our previous discussion on inter-ring exciton transfer. Long-time calculation for the case with an inter-ring distance of 15 $\rm \AA$ reveals that it takes about 200 phonon periods to have complete transfer of the exciton population, as shown in the inset of Fig.~\ref{Fig5}. As one phonon period is equal to about 20$fs$, the estimated total transfer time of 4$ps$ is in good agreement with the findings in Ref.~\cite{Hu2}.

Effects of static disorder on polaron dynamics have also been examined in order to test the applicability of the current approach to photosynthetic systems. Simulation results with Gaussian-distributed static disorder are included in Fig.~\ref{Fig5}. The on-site energy average is set to zero, and a standard deviation of $\sigma=0.1\omega_0$ is adopted in accordance with Ref.~\cite{silbey}. For simplicity, only on-site energy disorder is included although it will not be difficult to examine off-diagonal disorder. Over 100 realizations have been simulated in the calculation. Fig.~\ref{Fig5} provides a comparison on the exciton population transfer with and without the energetic disorder. While some population fluctuations have been smoothened out owing to disorder, the overall energy transfer efficiency is not significantly affected by the addition of static disorder. Furthermore, in the absence of any exact or more reliable dynamics studies, our approach here provide useful guide on the dynamics of excitonic energy transfer in LHI-LHII systems.

\subsection{Exciton Population Dynamics of N-site Symmetric Simple Ring Systems and LH2 Complexes}

Dynamics study in the previous section reveals the effect of dipolar interactions on exciton transfer efficiency at low temperatures, but it is also interesting to look at the opposite limit of high temperatures. As a starting point, it is preferable to use simple ring systems to analyze what may affect the energy transfer efficiency at high temperatures. Here we first study the energy transfer of symmetric ring systems of three, four, six, seven, eight and sixteen pigments. Efficiency of the energy transfer with respect to various trapping positions will be examined, and for an $N$-site ring system, analytical solutions will be obtained using the stationary condition on the off-diagonal density matrix element $\rho_{nm}$.

For a three-site ring system, we assume that all the site energies are the same, thus the energy difference is zero. The coupling between neighboring sites is assumed to be constant such that $J_{12}=J_{21}=J_{13}=J_{31}=J_{23}=J_{32}=J$. Furthermore, the coupling with the environment (dephasing) is  $\Gamma_{12}=\Gamma_{21}=\Gamma_{13}=\Gamma_{31}=\Gamma_{23}=\Gamma_{32}=\Gamma$. If the system is excited at site 1 and the excitation is trapped at the same location, the average trapping time is given by $\langle t \rangle={3}k_t^{-1}$, independent of $\Gamma$ and $J$. If a photon is absorbed at site 1, and the excitation is trapped at site 2, the average trapping time takes the form
\begin{equation}\label{eq8}
\langle t \rangle=\frac{3}{k_{t}}+\frac{1}{\Gamma}+\frac{1}{2\Gamma+k_{t}}+\frac{2\Gamma+k_{t}}{4J^{2}},
\end{equation}
which depends on $\Gamma$ and $J$. It is greater than that of the previous situation.

For a four-site ring system, if we only consider nearest neighbor couplings, we can easily get the analytical expression of the residence time and the average trapping  time. If a photon is absorbed at site 1, and the excitation is also trapped at site 1, the average trapping time is given by $\langle t \rangle={4}k_t^{-1}$. If a photon is absorbed at site 1, and the excitation is trapped at site 2 (or equivalently, site 4), the average trapping time is
\begin{equation}\label{eq10}
\langle t \rangle=\frac{4}{k_{t}}+\frac{1}{\Gamma}+\frac{1}{\Gamma+k_{t}/2}+\frac{3(2\Gamma+k_{t})}{8J^2}.
\end{equation}
If a photon is absorbed at site 1, and the excitation is trapped at site 3, the average trapping time is
\begin{equation}\label{eq11}
\langle t \rangle=\frac{4}{k_{t}}+\frac{\Gamma}{J^2}+\frac{3k_{t}}{8J^2}.
\end{equation}
By comparing the three equations for the four-site ring system, it is found that, the average trapping time of the first system is the smallest among the three scenarios, and that of the second is larger than that of the third.

As indicated by the definition of $\langle t \rangle$, exciton population is proportional to the value of average trapping time. Thus we can also probe the dynamics of the system using exciton population. It is interesting to note that the exciton population of site 3 grows larger than that of site 2 and site 4 even it is far from site 1, where the exciton is initially located, the phenomenon also happens for the second and third scenarios. To explain this, one can consider the two distinct paths for the exciton produced at site 1, in clockwise and and counterclockwise directions. They have the same phase when they meet at site 3, which makes the exciton population greatly enhanced due to quantum interference. As a result, the exciton population of site 3 is larger than that of site 2 and site 4 even though they are closer to site 1.

Furthermore, we note a few differences among these scenarios : $\rho_{1}(t)$ is larger in the second scenario than others, therefore, it has the largest average trapping time, which leads to the lowest energy transfer efficiency, and $\rho_{2}(t)$ and $\rho_{4}(t)$ are exactly the same in the first scenario due to positional symmetry. $\rho_{4}(t)$ is slightly larger in situation three than in situation one, because the trap at site 2 destroys the symmetry.

\subsection{Energy Transfer in Single Ring Systems}

\begin{figure}
\begin{center}
\includegraphics[height=!,width=8cm, bb= 90 137 535 470]{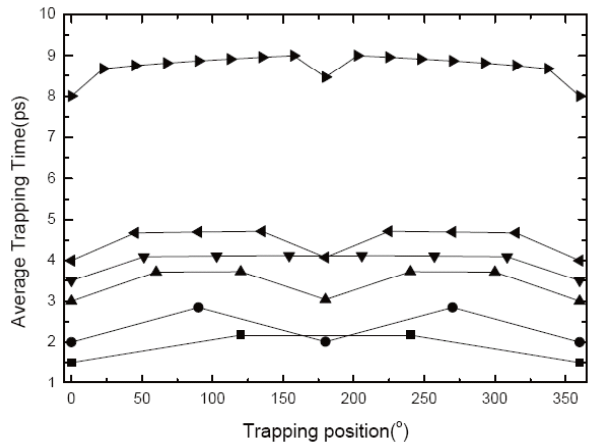}
\caption{The average trapping time versus varying trapping positions for simple single-ring systems. From bottom upwards, the square, circle, triangle, down triangle, left triangle and right triangle corresponds to three, four, six, seven, eight and sixteen-sites ring systems, respectively. For all systems, the parameters are chosen to $\hbar=1$, $k_{t}=2$, $\Gamma_{t}=20$, and $J=100$. }\label{Fig6}
\end{center}
\end{figure}


It is more interesting to look into the energy transfer efficiency of the ring system from the point of view of average trapping time, and as a starting point, simple single-ring systems with six, seven, eight and sixteen sites have been studied. The average trapping times at different trapping positions for the ring systems (with other parameters fixed to $\hbar=1, k_{t}=2, \Gamma_{t}=20, J=100$) are given in Fig.~\ref{Fig6}. Site 1 is located at the origin ($0^{\circ}$), and other sites are distributed in the ring uniformly. Taking a six-site system as an example, the zero-degree trapping corresponds to the case in which a photon is initially absorbed at site one and the excitation is trapped at the same site. Similarly, ($180^{\circ}$) trapping corresponds to the situation in which a photon is absorbed at site 1, and the excitation is trapped at site 4. It is interesting to find that the systems with an even number of sites have a comparatively smaller average trapping time when the exciton is trapped at the exactly opposite position (e.g., $180^{\circ}$) of photon absorption. Since this phenomenon does not occur in systems with odd number of sites, it can be inferred from the simple ring systems that the constructive and destructive interference, closely related to the symmetry of a system, plays a significant role in determining energy transfer efficiency.

For a ring system with even number of sites, it is possible to have constructive interference at the site exactly opposite to the photon absorption site in the ring, where the exciton amplitudes are now in phase due to their symmetrical transfer path. In this scenario, constructive interference provides almost exactly the same average trapping time compared to the case where exciton is created and trapped at the same site. However it should be noted that for systems with larger sizes, dephasing becomes increasingly important, resulting in a slight reduction in the energy transfer efficiency.

\begin{figure}
\begin{center}
\includegraphics[scale=0.26]{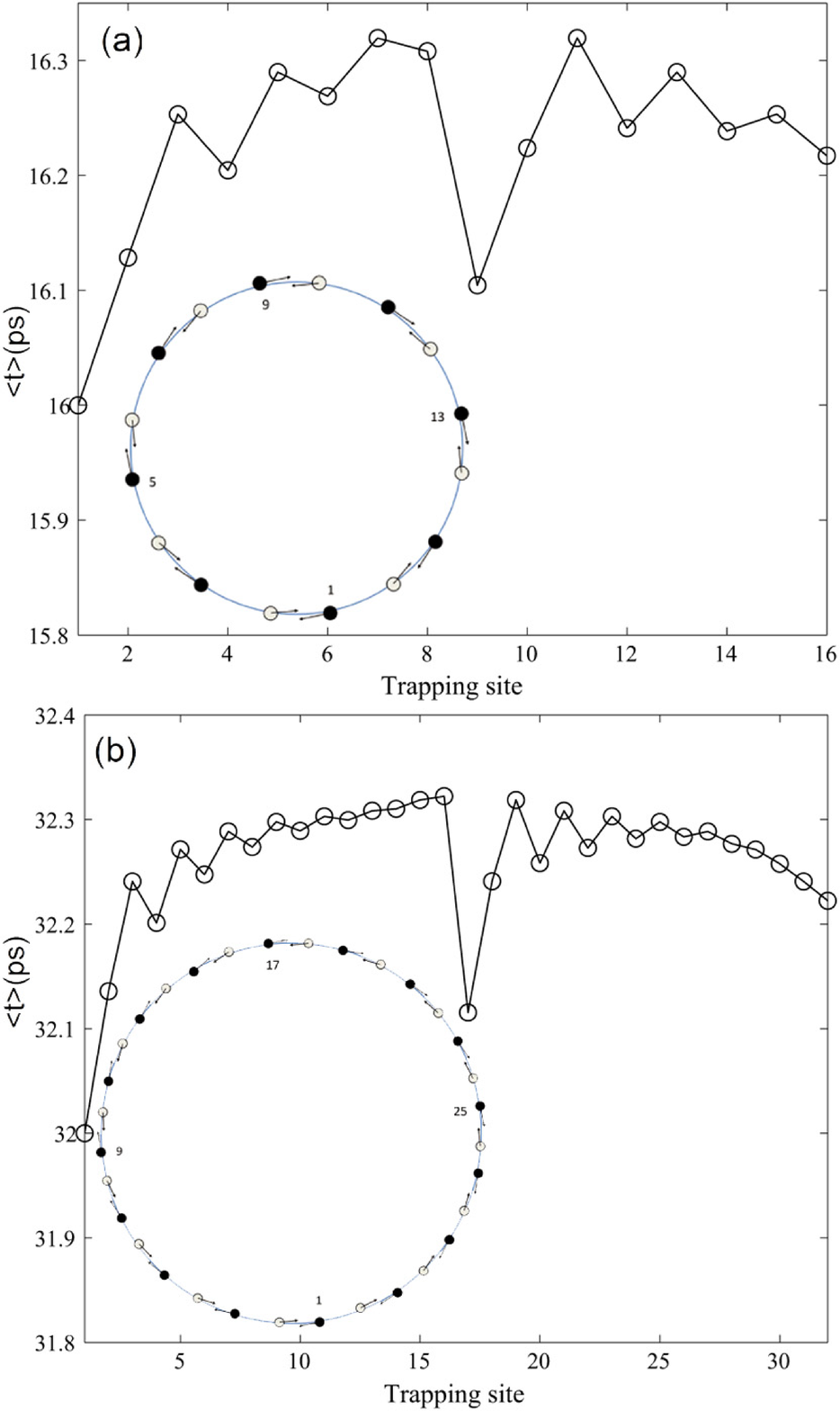}
\caption{Average trapping time versus trapping site for the inset schematics of B850 ring of (a) LH2, and (b) LH1. Solid dots represent for the
$\alpha$-apropoteins and hollow dots the $\beta$-apropoteins. The directions of the dipoles are tangential to the circle, with
$\alpha$-apropoteins pointing in clockwise direction, $\beta$-apropoteins anti-clockwise. A photon is absorbed at $1\alpha$ BChl, trapped at
different BChls. For all calculations, the trapping rate $k_t$ is set to 1 and dephasing rate $\Gamma$ is equal to 5. The rest of parameters, \emph{i.e., } $J_{nm}$, are given in Eq.~(\ref{J_nm}) and Eq.~(\ref{dipole-dipole}).}\label{Fig7}
\end{center}
\end{figure}

Now we turn to more realistic systems of photosynthesis, where capture of solar photons is often facilitated by light-harvesting antennae, and excitons, first created in the antennae, are transferred to reaction centers.

Motivated by the results of simple ring systems, we examine further how the trapping position would affect excitonic energy transfer in a realistic ring system that includes dipole-dipole interactions between all transition-dipole pairs that are not nearest neighbors. Fig.~\ref{Fig7}(a) displays the dependence of average trapping time on the trapping position, and the inset of Fig.~\ref{Fig7}(a), a schematic of B850 ring of LH2, in which solid dots denote the $\alpha$-apropotein, and the hollow dots, the $\beta$-apropotein. Interestingly, a dip corresponding to a minimal average trapping time is found to appear at site 9, in addition to the one at site 1. In comparison to the uniform ring systems shown in Fig.~\ref{Fig6}, there emerge zig-zag features in the trapping-time versus site-number plot in Fig.~\ref{Fig7}(a). Although the steepest dip can be simply explained with constructive interference, the appearance of multiple dips is directly related to the inclusion of long-range dipolar interactions, and therefore, multiple exction transfer pathways. Furthermore, it is also noted that the dependence of average trapping time on the site index no longer has the symmetries that exist for uniform rings. Furthermore, the non-symmetric dependence of $\langle t \rangle$ on the trapping position makes it possible to have different clockwise and anti-clockwise paths, an effect that can be related to the dimerization of the LH2 ring.

Dependence of average trapping time on trapping position has also been examined for a single LH1 ring, and the results are shown in Fig.~\ref{Fig7}(b). The exciton Hamiltonian of LH1 has a form similar to Eq.~(\ref{J_nm}) with the total number of sites increased from 16 to 32. Chromophores are assumed to be evenly distributed with a distance of about 10 ${\rm \AA}$, and other parameters for the LH1 system are taken from Ref.~\cite{Zhao}. Results similar to the LH2 case can be obtained from Fig.~\ref{Fig7}(b), and a dip appears around the middle point of the ring at site 17 with an obvious loss of symmetry. Due to the increase in the total site number, which mitigates interference effects, a much smoother curve is obtained in Fig.~\ref{Fig7}(b). For both the LH2 and LH1 rings, we have found that the long-range dipolar coupling reduces substantially the effect of quantum inference, in agreement with similar earlier findings in polaron dynamics simulation.

\subsection{Energy Transfer in Multiple Connected Ring Systems}

\begin{figure}
\begin{center}
\vspace{-0.0cm}
\includegraphics[scale=0.28]{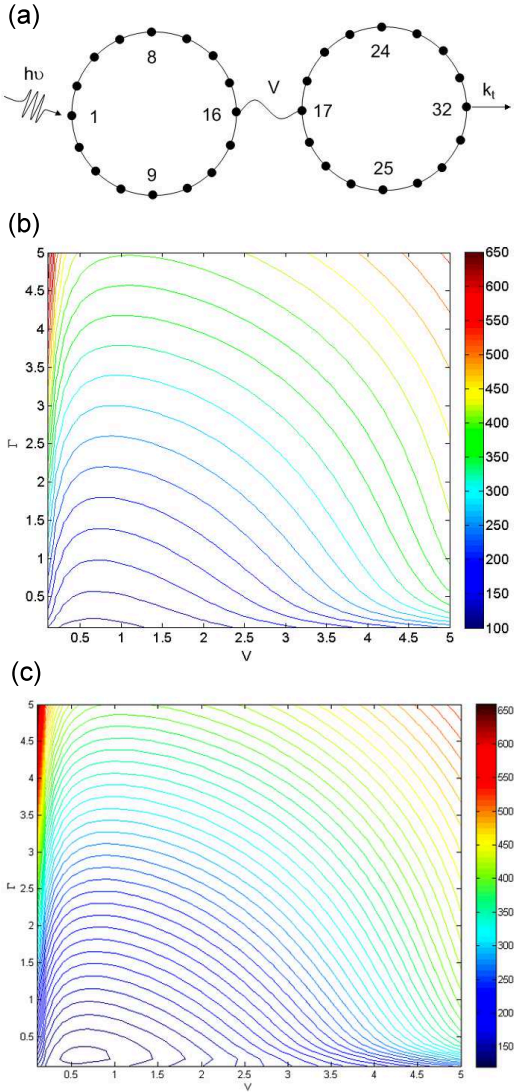}
\vspace{-0.0cm}
\caption{(a) Schematic of a two-ring toy model. (b) Contour map of $\langle t\rangle$ as a function of $\Gamma$ and V, with trapping site locating at site 32 of the second ring as shown in (a). (c) Contour map of $\langle t \rangle$ as a function of $\Gamma$ and V, with trapping site locating at site 31 of the second ring. The value of nearest neighbor coupling strength is set to J=1 and the trapping rate $k_t$ = 1.  }\label{Fig8}
\end{center}
\end{figure}

\begin{figure}
\begin{center}
\includegraphics[scale=0.35]{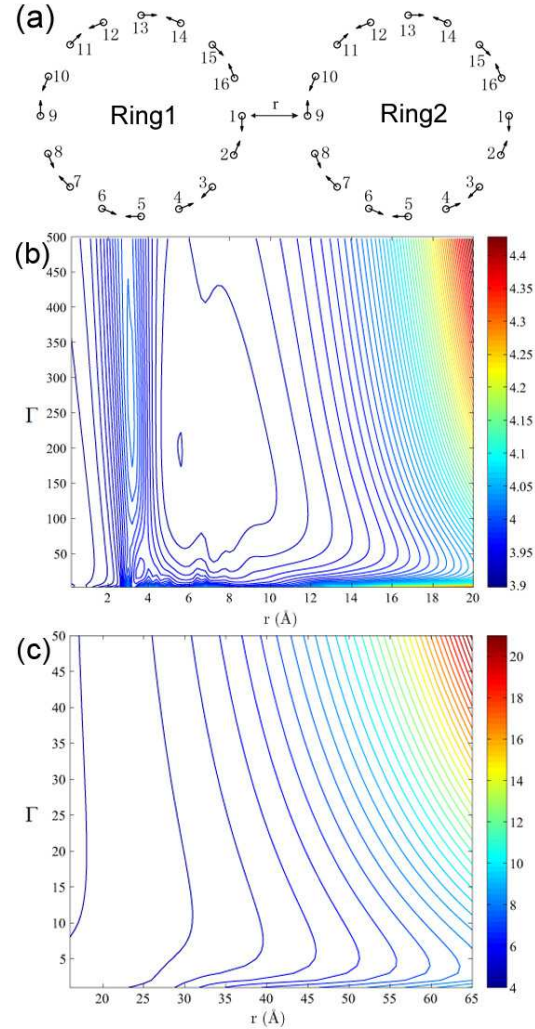}
\caption{(a) Schematic of a two-ring LH2 system. $r$ is the distance between the edges of the rings. (b) Short distance $\langle t \rangle$ contour
map as a function of $\Gamma$ and $r$. (c) Long distance $ \langle t \rangle$ contour map as a function of $\Gamma$ and $r$. In the calculations of $ \langle t \rangle$s, the first ring on the left is evenly excited, and exicton is evenly trapped in the ring-2. Inter-ring interaction takes same form of the intra-ring dipole-dipole coupling. The trapping rate in the calculations is assumed to be a constant, with $k_t$ = 1.  }\label{Fig9}
\end{center}
\end{figure}

To expand our study of energy transfer efficiency to multiple-ring systems, two-ring toy models as depicted in Fig.~\ref{Fig8} are examined first. Intra-ring nearest-neighbor interactions are assumed with coupling strength $J$, and site energies are set to be the same. The two rings are coupling through site 16 and site 17 with coupling strength $V$. Our findings are shown in Fig.~\ref{Fig8}, and an optimal coupling strength for efficient energy transfer is discovered at zero pure dephasing rate $\Gamma=0$. This optimal strength found is close to the value of intra-ring coupling $J$, a result of system symmetry loss. If the trapping site is moved from site 32 to site 31, the optimal trapping time is found at a nonzero pure dephrasing rate $\Gamma=0.2$, as shown in Fig.~\ref{Fig8}(c). It can be speculated that, to optimize energy transfer efficiency, system symmetry imperfections can be compensated by the introduction of pure dephasing or dissipation of system coherence by the environment. Such a finding is applicable for design of artificial photosynthetic systems in general.

Next we will look at a realistic two-ring LH2 system with long-range dipolar interactions as illustrated in Fig.~\ref{Fig9}(a). Two rings interact via dipole-dipole coupling, according to Eq.~(\ref{dipole-dipole}). An initial state in which all sites in the first ring are evenly excited is selected, \emph{i.e.,} at $t=0$, $\rho_{nn}=1/16$ for $n=1, 2, ..... 16$ in the first ring. We also assume that the excitation is evenly trapped in the second ring. It is found that the trapping time increases with the edge-to-edge interring distance $r$ as expected, and for $ r>20{\rm \AA}$ (or with higher dephasing rates), the trapping time increases rapidly. It is known that quantum coherence may help enhance the efficiency of energy transfer \cite{Alexandra}. At higher dephasing rates, with the destruction of coherence, the excitation is transported classically, resulting in longer trapping times and lower efficiency.

The findings on average trapping time in Fig.~\ref{Fig9} suggest extremely distance-dependent nature of energy transfer. As indicated in Ref.~\cite{nchem}, the sites too close to each other in certain orientations will lead to non-fluorescent dimers, which can serve as excitation sinks that deplete excitations.

One can also observe an optimal value of inter-ring distance of around $\rm {5 \AA}$, corresponding to the highest strength of dipolar interaction between the nearest neighbor dipoles. This value is not close to the realistic inter-ring distance of LH2 complexes in a close packed membrane. Although it has been pointed out in Ref.~\cite{nchem} that a typical inter-chromophore center-to-center distances of neighboring molecules are consequently close (~6 to 25 $\AA$). Thus, the global minimum obtained with the method introduced in this paper is only an ideal case such that two molecular aggregates are in contact with each other without any support of
protein scaffold. Although the value only corresponds to the ideal case with highest inter-ring coupling strength, the existence of the local minimum is non-negligible. Such local minimum in average trapping time with non-zero dephasing rate appears in a continuous fashion throughout a large range of distances between the rings as shown in Fig.~\ref{Fig9}(b) and (c). The existence of this local minimum can be explained with the role of environment related dephasing in increasing energy transfer efficiency for a system with asymmetric properties. The symmetry-breaking dipolar interaction in the LH2 system leads to multiple non-nearest neighbor energy transfer pathways, which can enhance the energy transfer within the ring greatly as shown in the polaron dynamics calculations, however it is more preferable to have more exciton populations that can be trapped at certain site. Thus it is preferable to have finite dephasing to increase the inter-ring transfer efficiency by maximizing the amount of exciton populations that can be trapped in the trapping site of the second ring through multiple pathway interference. This small amount of dephasing is not sufficient to destroy the quantum coherence in the system, which gives us a hint on how the unique arrangements and interactions between chromophores in the LH2 complexes as well as the way that each complex interacts with others in optimizing the energy transfer in the presence of an environment that is coupled to the system. More interestingly, in a recent work on Fenna-Matthews-Olson (FMO) complex \cite{PRE}, even for this finite-size, disordered molecular network can effectively preserve coherent excitation energy transfer against ambient dephasing, which lends a further support to what we have revealed in this section. Furthermore, the role of dipolar interaction in increasing exciton transfer efficiency through introducing of multiple pathway for inter-ring exciton transfer is evident, which further supports the previous findings.

\section{Conclusions}

In this paper, two distinct approaches have been applied to study the excitation energy transfer in the ring systems. From both methods, non-nearest neighbor dipolar interaction is found to be helpful in increasing intra- or inter-ring exciton transfer efficiency as a result of multiple pathways.

Using the Haken-Strobl model, the energy transfer in the ring systems with coupling to environment is studied in details. The energy trapping efficiency is found to be dependent on the trapping position in both the hypothetic and the realistic ring systems. For the hypothetic system, with the trap positioned diametrically opposite to the site of the initial excitation (which is possible only in systems with even number of sites), we observe a sudden drop of the trapping time due to the constructive interference between the clockwise and anti-clockwise paths leading to the trap. The situations in a LH2 (B850) ring and LH1 ring are similar to the hypothetic systems, but owing to the broken symmetry caused by the dimerization of BChls and dipolar couplings, the drop of the trapping time is not significant. In other words, the ring becomes more homogeneous in terms of energy transfer, which further supports the findings of polaron dynamics studies of one-ring system.

The polaron dynamics of the coupled rings revealed interesting role of dipolar interaction in increasing energy transfer efficiency by introducing multiple transfer paths between the rings, and the time scale of inter-ring exciton transfer is in good agreement with previous studies. Moreover, when the Haken-Strobl model is applied for the coupled rings, an optimal coupling strength is obtained between the hypothetic two-ring systems with zero dephasing rate, while in a two-LH2 system, owing to the intrinsic symmetry breaking, global minimum and local minimum of average trapping time is always found with non-zero depahsing rate. In both the cases, the efficiency drops with further increase of dephasing rate, due to the loss of quantum coherence. The findings in this paper regarding exciton transfer processes in ring systems might assist us to design more efficient artificial light harvesting systems.

\section*{Acknowledgments}

Support from the Singapore National Research Foundation through the Competitive Research  Programme (CRP) under Project No.~NRF-CRP5-2009-04 and the Singapore Ministry of Education through the Academic Research Fund (Tier 2) under Project No.~T207B1214 is gratefully acknowledged.

\appendix
\section{Precision of the Time-dependent Davydov Ans\"{a}tze}

For a trial wave function $|\Psi(t)\rangle$ that does not strictly obey the time-dependent Schr\"{o}dinger equation, the deviation vector
$|\delta(t)\rangle$ can be defined as
\begin{equation}\label{DetlaVector}
|\delta(t)\rangle \equiv i\frac{\partial}{\partial t}|\Psi(t)\rangle - \hat{H}|\Psi(t)\rangle,
\end{equation}
and the deviation amplitude $\Delta(t)$ is defined as
\begin{equation}\label{deltadef2}
\Delta(t) \equiv \sqrt{\langle\delta(t)|\delta(t)\rangle}.
\end{equation}

For the Holstein Hamiltonian $\hat{H}$ defined in Eqs.~(\ref{H_tot})-(\ref{H_diag}), one can derive the explicit expression of
$\langle\delta(t)|\delta(t)\rangle$ for the ${\rm D}_1$ Ansatz:
\begin{eqnarray}\label{delta2D1}
&&\Big\langle\delta_{{\rm D}_1}(t)\Big|\delta_{{\rm D}_1}(t)\Big\rangle \nonumber\\
&=& \sum_n\Big|i\dot{\alpha}_n(t) + T_n(t) + \alpha_n(t)R_n(t)\Big|^2 \nonumber\\
&+& \sum_{\alpha_n(t)\neq0}\sum_q\Big|\alpha_n(t)\big[i\dot\lambda_{n,q}(t) + \frac{g_q}{\sqrt{N}}\omega_qe^{-iqn} - \omega_q\lambda_{n,q}(t)\big] \nonumber\\
&&+ \Omega_{n,q}(t)\Big|^2 + \Delta_{\rm D}^2(t) - \sum_{\alpha_n(t)\neq0}\sum_q\Big|\Omega_{n,q}(t)\Big|^2
\end{eqnarray}
with
\begin{equation}\label{RnD}
R_n(t) \equiv {\rm Re}\sum_q\big[i\dot\lambda_{n,q}(t) + \frac{2g_q}{\sqrt{N}}\omega_qe^{-iqn} - \omega_q\lambda_{n,q}(t)\big]\lambda_{n,q}^*(t),
\end{equation}
where $T_n(t)$, $\Omega_{n,q}(t)$ and $\Delta_{\rm D}^2(t)$ are three expressions which have no item of $\dot{\alpha}_n(t)$ or
$\dot\lambda_{n,q}(t)$.

For the ${\rm D}_2$ Ansatz, Eq.~(\ref{delta2D1}) can be further derived to
\begin{eqnarray}\label{delta2D2}
&&\Big\langle\delta_{{\rm D}_2}(t)\Big|\delta_{{\rm D}_2}(t)\Big\rangle \nonumber\\
&=& \sum_n\Big|i\dot{\alpha}_n(t) + T_n(t) + \alpha_n(t)R_n(t)\Big|^2 \nonumber\\
&+& \sum_q\Big|i\dot{\beta}_q(t) + \frac{g_q}{\sqrt{N}}\omega_q\sum_n |\alpha_n(t)|^2e^{-iqn} - \omega_q\beta_q(t)\Big|^2 \nonumber\\
&+& \frac{1}{N}\sum_q g_q^2\omega_q^2\Bigg[1 - \Big|\sum_n |\alpha_n(t)|^2e^{-iqn}\Big|^2\Bigg].
\end{eqnarray}
And for the $\tilde{\rm D}$ Ansatz, Eq.~(\ref{delta2D1}) can be further derived to
\begin{eqnarray}\label{delta2Dtilde}
&&\Big\langle\delta_{\tilde{\rm D}}(t)\Big|\delta_{\tilde{\rm D}}(t)\Big\rangle \nonumber\\
&=& \sum_n\Big|i\dot{\alpha}_n(t) + T_n(t) + \alpha_n(t)R_n(t)\Big|^2 \nonumber\\
&+& \Big|i\dot{\lambda}_0(t) + \frac{g_q}{\sqrt{N}}\omega_0 - \omega_0\lambda_0(t)\Big|^2 \nonumber\\
&+& \sum_{q\neq 0} \Big|i\dot{\gamma}_q(t) + \xi_q(t)i\dot{\beta}_q(t) + y_q(t)\Big|^2 \nonumber\\
&+& \sum_{q\neq 0} \Big[1-|\xi_q(t)|^2\Big]\left|i\dot{\beta}_q(t) + \frac{d_q(t)}{1-|\xi_q(t)|^2}\right|^2 \nonumber\\
&+& \sum_{q\neq 0}\left\{\sum_n\Big|\Theta_{n,q}(t)\Big|^2 - \left[|y_q(t)|^2+\frac{|d_q(t)|^2}{1-|\xi_q(t)|^2}\right]\right\} \nonumber\\
&+& \Delta_{\rm D}^2(t) - \sum_{n}\sum_q\Big|\Omega_{n,q}(t)\Big|^2,
\end{eqnarray}
where $\xi_q(t)$, $\Theta_{n,q}(t)$, $y_q(t)$ and $d_q(t)$ are four expressions which have no item of $\dot{\alpha}_n(t)$, $\dot\beta_{q}(t)$,
$\dot\gamma_{q}(t)$ or $\dot\lambda_0(t)$.

Substituting Eq.~(\ref{localDavyAnsatz}) into Eqs.~(\ref{H_ex})-(\ref{H_diag}), one obtains the expressions for the system energies by the Davydov
Ans\"{a}tze in the Holstein model:
\begin{eqnarray}\label{Eex}
E_{\rm ex}(t) &\equiv& \big\langle\Psi_{\rm D}(t)\big|\hat{H}_{\rm ex}\big|\Psi_{\rm D}(t)\big\rangle \nonumber \\
&=& -2J{\rm Re}\sum_{n}\alpha_{n}^{*}(t)S_{n,n+1}(t)\alpha_{n+1}(t),
\end{eqnarray}
\begin{eqnarray}\label{Eph}
E_{\rm ph}(t) &\equiv& \big\langle\Psi_{\rm D}(t)\big|\hat{H}_{\rm ph}\big|\Psi_{\rm D}(t)\big\rangle \nonumber \\
&=& \sum_{n}\Big[|\alpha_{n}(t)|^{2}\sum_{q}\omega_{q}|\lambda_{n,q}(t)|^{2}\Big],
\end{eqnarray}
and
\begin{eqnarray}\label{Ediag}
E_{\rm ex-ph}(t)&\equiv&\big\langle\Psi_{\rm D}(t)\big|\hat{H}_{\rm ex-ph}\big|\Psi_{\rm D}(t)\big\rangle\nonumber \\
&=& -\frac{2}{\sqrt{N}}\sum_{n}\Big[|\alpha_{n}(t)|^{2}{\rm Re}\sum_{q}g_q\omega_{q}\lambda_{n,q}(t)e^{iqn}\Big], \nonumber \\
\end{eqnarray}
where $S_{n,m}(t)$ is the Debye-Waller factor.

Note that since the unit of $\Delta(t)$ is that of the energy, by comparing $\Delta(t)$ with the main component of the system energies such as $E_{\rm ph}(t)$ and $E_{\rm ex-ph}(t)$, one can observe whether the deviation of an Ansatz from obeying the Schr\"{o}dinger equation is negligible or not, in the concerned case. From this perspective, the comparison between $\Delta(t)$ and energy components of the system provides a good reference for the validity of an Ansatz.\\

\section{Detail derivation of polaron dynamics in multiple-ring system}

The system energies of the multiple-ring system can be obtained as follows by apply Davydov Ans\"{a}tze to the modified Holstein Hamiltonian given in Eq.~(\ref{H_ex2}), (\ref{H_ph2}) and (\ref{H_exph2}):
\begin{equation}
E_{\rm ex}(t)=-\sum_{r_1 r_2}\sum_{nm}J^{r_1 r_2}_{nm}\alpha_n^{r_1 \ast}(t)\alpha_m^{r_2}(t)S_{nm}^{r_1 r_2}(t)
\end{equation}
\begin{equation}
E_{\rm ph}(t)=\sum_r \sum_n \vert \alpha_n^r(t) \vert ^2 \sum_q \omega_q^r \vert \lambda_{nq}^r(t) \vert ^2
\end{equation}
\begin{equation}
E_{\rm ex-ph}(t)=-\frac{2}{\sqrt{N}}\sum_r \sum_n \sum_q g^r_q\omega_q^r \vert \alpha_n^r(t) \vert ^2 {\rm Re}[e^{iqn}\lambda_{nq}^r(t)]
\end{equation}
with Debye-Waller factor $S_{nm}^{r_1 r_2}(t)$ given by:
\begin{eqnarray}
S^{r_1 r_2}_{nm}(t) &=& {\rm exp}\{\sum_q[\lambda^{r_1\ast}_{nq}(t)\lambda^{r_2}_{mq}(t)\delta_{r_1 r_2} \nonumber\\
&-&\frac{1}{2}\vert\lambda^{r_1}_{nq}(t)\vert^2-\frac{1}{2}\vert\lambda^{r_2}_{mq}(t)\vert^2]\},
\end{eqnarray}
and the index $r_1$ and $r_2$ all runs over 1 to $N_{\rm ring}$.

Since D1 is the used for all calculations in this paper, here we only give the expression of $\langle\delta(t)|\delta(t)\rangle$ for D1, which can be derived as follows:
\begin{eqnarray}\label{dela2D1rings}
&&\langle\delta_{D_1}(t)\vert\delta_{D_1}(t)\rangle \nonumber\\
&=&\sum_r \sum_n\vert i\dot{\alpha}_n^r(t)+\alpha_n^r(t)R_n^r(t)+T_n^r(t)\vert^2\nonumber\\
&+&\sum_r\sum_{\alpha^r_n(t)\not=0}\sum_q\vert\alpha^r_n(t)\varpi^r_{nq}(t)+\Omega^r_{nq}(t)\vert^2+\Delta^2_D(t)
\nonumber\\
&-&\sum_r\sum_{\alpha^r_n(t)\not=0}\sum_q\vert\Omega^r_{nq}(t)\vert^2-\sum_r\sum_n\vert T^r_n(t)\vert^2.
\end{eqnarray}

Minimization of the first two terms in Eq.~(\ref{dela2D1rings}) lead to the equations of motions for the time-dependent variational parameters $\alpha^r_n(t)$ and $\lambda^r_{nq}(t)$ as:
\begin{equation}\label{alphadt2}
\dot{\alpha}^r_n(t)=i[T^r_n(t)+\alpha^r_n(t)R^r_n(t)],
\end{equation}
and
\begin{equation}\label{lambdadt2}
\dot{\lambda}^r_{nq}(t)=i[\frac{\Omega^r_{nq}(t)}{\alpha^r_n(t)}+\frac{g^r_q}{\sqrt{N}}\omega^r_q e^{-iqn}-\omega^r_q\lambda^r_{nq}(t)].
\end{equation}

Each collected term in Eqs.~(\ref{alphadt2}) and~(\ref{lambdadt2}) is given as:
\begin{equation}
\varpi^r_{nq}(t)=i\dot{\lambda}^r_{nq}(t)+\frac{g^r_q}{\sqrt{N}}\omega^r_q e^{-iqn}-\omega^r_q\lambda^r_{nq}(t),
\end{equation}
\begin{equation}
R^r_n(t)=Re\sum_q[\varpi^r_{nq}(t)+\frac{g^r_q}{\sqrt{N}}\omega^r_q e^{-iqn}]\lambda^{r\ast}_{nq}(t),
\end{equation}
\begin{equation}
T^{r_1}_n(t)=\sum_{r_2}\sum_m j^{r_1 r_2}_{nm}\alpha^{r_2}_m(t) S^{r_1 r_2}_{nm}(t),
\end{equation}
\begin{equation}
\Omega^{r_1}_{nq}(t)=\sum_{r_2}\sum_m j^{r_1 r_2}_{nm}\alpha^{r_2}_m(t) S^{r_1 r_2}_{nm}(t)[\lambda^{r_2}_{mq}(t)\delta_{r_1
r_2}-\lambda^{r_1}_{nq}(t)],
\end{equation}
and
\begin{equation}
\Delta^2_{\rm D}(t)=\sum_{r_1}\sum_n[\sum_{r_2'm'}\sum_{r_2 m}j^{r_1 r_2'}_{nm'}j^{r_1 r_2}_{nm}\alpha^{r_2'*}_{m'}(t)\alpha^{r_2}_{m}(t)S^{r_2'
r_2}_{m' m}(t)].
\end{equation}


\begin{thebibliography}{99}

\bibitem{Sauer} K.~Sauer, Annu.~Rev.~Phys.~Chem., {\bf 30}, 155, (1979),

\bibitem{Blankenship} R.~E.~Blankenship, \textit{Molecular Mechanisms of photosynthesis.}; Blackwell Science: Oxford/Malden, 2002.

\bibitem{Grondelle} R.~van Grondelle, V.~I.~Novoderezhkin., Phys.~Chem.~Chem.~Phys., {\bf 8}, 793, (2006).


(2004).

\bibitem{Brixner} T.~Brixner, J.~Stenger, H.~M.~Vaswani, M.~Cho, R.~E.~Blankenship, G.~R.~Fleming, Nature(London),  {\bf 434}, 625, (2005).


\bibitem{Scholes} G.~D.~Scholes, Annu.~Rev.~Phys.~Chem., {\bf 54}, 57, (2003).

\bibitem{Cheng} Y.~C.~Cheng, G.~R.~Fleming, Annu.~Rev.~Phys.~Chem. {\bf 60}, 241, (2009).

\bibitem{Jang1} S.~Jang, M.~D.~Newton, and R.~J.~Silbey, Phys.~Rev.~Lett. {\bf 92}, 218301, (2004).

\bibitem{Jang2} S.~Jang, M.~D.~Netwton, and R.~J.~Silbey, J.~Phys.~Chem.~B {\bf 111}, 6807, (2007).

\bibitem{Jang3} S.~Jang, J.~Chem.~Phys. {\bf 127}, 174710 (2007).

\bibitem{Jang4} S.~Jang, M.~D.~Newton, and R.~J.~Silbey, J.~Chem.~Phys. {\bf 118}, 9312 (2003).

\bibitem{Jang5} S.~Jang and R.~J.~Silbey, J.~Chem.~Phys. {\bf 118}, 9324 (2003).

\bibitem{Breuer} H.~P.~Breuer, F.~Petruccione, \textit{The Theory of Open Quantum Systems.} Oxford University Press: Oxford, 2002.

\bibitem{Redfield} A.G.~Redfield,  IBM J.~Res.~Dev., {\bf 1}, 19, (1957).

\bibitem{Itzykson} Itzykson and J. Zuber, \emph{Many-Particle Physics}, McGraw-Hill, New York, (1980).

\bibitem{Knoester} D.~J.~Heijs, V.~A.~Malyshev, and J.~Knoester, Phys.~Rev.~Lett. {\bf 95}, 177402 (2005).

\bibitem{Schreiber} M.~Schr$\ddot{o}$der, U.~Kleinekathofer, and M.~Schreiber, J.~Chem.~Phys. {\bf 124}, 084903 (2006).

\bibitem{Coker} P.~Huo and D.~F.~Coker, J.~Chem.~Phys. {\bf 133}, 184108, (2010).

\bibitem{Haken} H.~Haken, G.~Strobl, Z.~Phys., {\bf 262}, 135, (1973).

\bibitem{Hu1} X.~Hu, T.~Ritz, A.~Damjanovi\'c, F.~Autenrieth, K.~Schulten, Q.~Rev.~Biophys., {\bf 35}, 1, (2002).

\bibitem{McDermott} G.~McDermontt, S.~M.~Prince, A.~A.~Freer, A.~M.~Hawthornthwaite, M.~Z.~Papiz, R.~J.~Cogdell, N.~W.~Issacs, Nature(London), {\bf 374}, 517, (1995).

\bibitem{Koepke} J.~Koepke, X.~Hu, C.~Muenke, K.~Schulten, H.~Michel, Structure {\bf 4}, 581, (1996).

\bibitem{Fenna} R.~E.~Fenna, B.~W.~Matthews, J.~M.~Olson, E.~K.~Shaw, J.~Mol.~Biol., {\bf 84}, 231, (1974).

\bibitem{Zhao} Y.~Zhao, M.~F.~Ng, G.~H.~Chen, Phys.~Rev.~E, {\bf 69}, 032902, (2004).

\bibitem{Alexandra} A.~Olaya-castro, C.~F.~Lee, F.~F.~Olsen, N.~F.~Johnson, Phys.~Rev.~B, {\bf 79}, 065115, (2008).

\bibitem{Fassioli} F.~Fassioli, A.~Olaya-castro, S.~Scheuring, J.~N.~Sturgis, N.~F.~Johnson, Biophys.~J., {\bf 97}, 2464, (2009).

\bibitem{Muh} F.~M\"uh, M.~EI-A.~Madjet, J.~Adolphs, A.~Abdurahman, B.~Rabenstein, H.~Ishikita, E.~W.~Knapp, T.~Renger, Proc.~Natl.~Acad.~Sci.~USA, {\bf 104}, 16862, (2007).

\bibitem{Ishizaki} A.~Ishizaki, G.~R.~Fleming, Proc.~Natl.~Acad.~Sci.~USA, {\bf 106}, 17255, (2009).

\bibitem{Cao1} J.~S.~Cao, R.~J.~Silbey, J.~Phys.~Chem.~A, {\bf 113}, 13825, (2009).

\bibitem{Cao2} J.~H.~Kim, J.~S.~Cao, J.~Phys.~Chem.~B, {\bf 114}, 16189, (2010).

\bibitem{Holstein1959} T.~Holstein, Ann. Phys., {\bf 8}, 325 (1959); {\it ibid}, {\bf 8}, 343, (1959).

\bibitem{Mahan} G.D.~Mahan, {\it Many Particle Physics}, 3rd edition (Kluwer Academic, 2000).

\bibitem{Ulrich} C.~Olbrich and U.~Kleinekath\"ofer, J.~Phys.~Chem.~B {\bf 114}, 12427, (2010).

\bibitem{Hofmanm} C.~Hofmann, T.~J.~Aartsma, H.~Michel, J.~K\"ohler, Proc.~Natl.~Acad.~Sci.~USA, {\bf 100}, 15534, (2003).

\bibitem{Schulten1} A.~Damjanov\'c, I.~Kosztin, U.~Kleinekath\"ofer, and K.~Schulten, Phys.~Rev.~E {\bf 65}, 031919, (2002).

\bibitem{Cao3} J.~Cao, R.~J.~Silbey, J.~Phys.~Chem. {\bf 112}, 12867, (2008).

\bibitem{Hu2} X.~Hu, T.~Ritz, A.~Damjanovi\'c, K.~Schulten, J.~Phys.~Chem.~B, {\bf 101}, 3854, (1997).

\bibitem{pccp} B.~Luo, J.~Ye, C.B.~Guan and Y.~Zhao, Phys.~Chem.~Chem.~Phys., {\bf 12}, 15073, (2010).

\bibitem{pssc} B.~Luo, J.~Ye, and Y.~Zhao, Phys. Phys.~Status~Solidi~C, {\bf 8}, 70, (2011).

\bibitem{silbey} S.~Jang, S.~E.~Dempster, and R.~J.~Silbey, J.~Phys.~Chem.~B, {\bf 105}, 6655, (2001).

\bibitem{absexp} J.~L.~Herek \emph{et al.}, Biophys.~J., {\bf 78}, 2590, (2000).

\bibitem{nchem} G.~D.~Scholes \emph{et al.}, Nature Chem., {\bf 3}, 763, (2011).

\bibitem{PRE} T.~Scholak \emph{et al.}, Phys.~Rev.~E, {\bf 83}, 021912, (2011).

\end{thebibliography}
\end{document}